\documentclass{aa}
\usepackage{graphicx}
\usepackage{natbib}
\usepackage{txfonts}
\usepackage{epstopdf}
\bibpunct{(}{)}{;}{a}{}{,}    

\def\apjs{Astrophys. J. Supp.}
        
\def\jgr{J. Geophys. Res.}
          
\def\grl{Geophys. Res. Lett.}

\def\aap{Astron. Astrophys.}

\def\apj{Astrophys. J.}
 
 \def\apjl{Astrophys. J. Lett.}         
                   
\def\solphys{Sol. Phys.}
     
\def\nat{Nature}

\def\aj{Astron. J.}

\begin{document}

\title{The role of the Fraunhofer lines in solar brightness variability}
\author{A.I. Shapiro \inst{1} \and S.K. Solanki \inst{1,2} \and N.A. Krivova \inst{1} \and R.V. Tagirov \inst{3,4} \and W.K. Schmutz \inst{3}  }  
\offprints{A.I. Shapiro}

\institute{Max-Planck-Institut f{\"u}r Sonnensystemforschung, Justus-von-Liebig-Weg 3, 37077, G{\"o}ttingen, Germany\\
\email{shapiroa@mps.mpg.de}
\and School of Space Research, Kyung Hee University, Yongin, Gyeonggi 446-701, Korea
\and Physikalisch-Meteorologishes Observatorium Davos, World Radiation Centre, 7260 Davos Dorf, Switzerland
\and Institute of Astronomy, ETH Zentrum, CH-8092 Zurich, Switzerland\\}

\date{Received ; accepted }

\abstract
{The solar brightness varies on timescales from minutes to decades. A clear identification of the physical processes behind such variations is needed  for developing and improving  physics-based models of solar brightness variability and reconstructing solar brightness in the past.
This is, in turn,  important for better understanding the solar-terrestrial and solar-stellar connections.}
{We estimate the relative contributions of the continuum, molecular, and atomic lines to the solar brightness variations on different timescales.}
{Our approach is based on the assumption that variability of the solar brightness on timescales greater than a day is driven by the evolution of the solar surface magnetic field. We calculated the solar brightness variations employing the solar disc area coverage of magnetic features deduced from the MDI/SOHO observations. The brightness contrasts of magnetic features relative to the quiet Sun were calculated with a non-LTE radiative transfer code as functions of disc position and wavelength. By consecutive elimination of molecular and atomic lines from the radiative transfer calculations, we assessed the role of these lines in producing solar brightness variability.} 
{We show that the variations in Fraunhofer lines define the amplitude of the solar brightness variability on timescales greater than a day and even the phase of the total solar irradiance variability over the 11-year cycle. We also demonstrate that molecular lines make substantial contribution to solar brightness variability on the 11-year activity cycle and centennial timescales. In particular, our model indicates that roughly a quarter of the total solar irradiance variability over the 11-year cycle originates in molecular lines.  
The maximum of  the absolute spectral brightness variability  on timescales greater than a day is associated with the CN violet system between 380 and 390 nm.}
{}


\titlerunning{The role of the Fraunhofer lines in solar brightness variability}
\maketitle

\section{Introduction}\label{sect:intro}
Regular spaceborne  measurements of the total solar irradiance (TSI, which is the spectrally integrated solar radiative flux at one au from the Sun) started in 1978 with the launch of the NIMBUS 7 mission \citep{hoytetal1992}. They revealed that the TSI varies on multiple timescales \citep{wilson1981, frohlich2005}. The most striking features of the TSI records are the $\sim$ 0.1\% modulation over the course of the 11-year solar activity cycle and more irregular variations on the timescale of solar rotation with {   amplitudes} of up to 0.3\%.

Measurements of the spectral solar irradiance (SSI, which is the solar radiative flux per unit wavelength at one au from the Sun) showed the variability on the same two timescales  \citep[see][and references therein]{Floydetal2003,harderetal2009, delandandcebula2012}. They also indicated  that the amplitude of the {\it relative} SSI variations strongly depends on the wavelength and  generally increases towards short wavelengths. For example,  the variability of the solar Ly-$\alpha$ line (121.5 nm) irradiance over the course of the 11-year cycle reaches 100\% \citep{woodsetal2000};  i.e., it is three {   orders} of magnitude higher than the TSI variability.

It is widely accepted now that  the SSI variability on timescales of a day and longer is caused by solar surface magnetic fields  \citep[][]{domingoetal2009,krivovaetal_cycle23,TOSCA2012,MPS_AA,yeo_rev}, although several other physical mechanisms for SSI variability have been proposed \citep[see e.g.][]{kuhnandstein1996,kuhnetal1998}.  

Magnetic fields emerge on the solar surface in the form of magnetic concentrations described well by flux tubes  \citep[see][for reviews]{solanki_FT, Sami_B}. Large flux tubes usually form dark sunspots, while small flux tubes appear as bright features ensembles of which compose faculae and the network, while individual small flux tubes are seen as bright points when observed at high resolution.  The effects from the dark and bright features on solar irradiance do not compensate each other and the imbalance depends on the solar disc coverage by magnetic features of different sizes {   and on the wavelength observed}. The coverage changes with time, leading to the variability of solar irradiance on timescales from days to decades. 
In recent years a number of models of solar irradiance variability  have been created  that are to some {   extent} based on such a concept \citep[e.g.][]{fliggeetal2000, krivovaetal2003, leanetal2005, shapiro_rec, fontenlaetal2011, bolducetal2014,balletal2014, yeoetal2014}.

The interest in solar irradiance variability is not limited to the solar community. {   The terrestrial climate} responds to the decadal variations in solar irradiance \citep[see e.g. reviews by][]{haigh2007,grayetal2010,TOSCA2012,MPS_AA}  and there is also evidence for a longer term influence of solar activity on climate \citep[e.g.][]{bondetal2001}. Studies of solar irradiance are also of high importance for stellar astronomers, who have been comparing it with the variability of other lower main sequence stars \citep{radicketal1998,lockwoodetal2007,halletal2009}. The interest in solar-stellar comparisons has recently been rekindled \citep[see e.g.][]{McQuillanetal2012,basrietal2013} by the unprecedented precision of broadband stellar photometry achieved with the launch of the Kepler \citep{KEPLER}  and Corot \citep{COROT2,COROT} space missions and the anticipation of the upcoming PLATO mission \citep{PLATO}.


Changes in the solar spectrum can be decomposed into the variations in the continuum and in spectral lines. The spectral lines are present all over the solar spectrum and they are responsible for its extremely rich and sophisticated structure.
In the EUV {   (10--100 nm) and far UV (100--200 nm) below 170 nm} the solar spectrum contains strong emission lines,  while longward of 170 nm its profile is determined by many millions  of atomic and molecular absorption lines, also called Fraunhofer lines. It is hardly possible to find a spectral interval without Fraunhofer lines in the visible and near-infrared parts of the solar spectrum. In the UV the immense number of spectrally unresolved lines form the UV {\it line haze} which completely blocks the continuum photons \citep[e.g.][]{collet2005,shorthauschildt2005}.

The variations in the Fraunhofer lines in the full-disc solar brightness spectrum over the course of the activity cycle were first revealed in the pioneering studies by \cite{lines1} who analysed the Kitt Peak spectral records for seven Fraunhofer lines and found a decrease of equivalent width ranging from no variation {   (Si I 10827.1 {\AA} line) to 2.3\% (C I 5380.3  {\AA} line) }from  1976 to 1980. \cite{lines1}  suggested that this decrease may be attributed to a change in the photospheric temperature gradient \citep[see also][]{lines3}. Later \cite{mitchellandlivingston1991} employed Fourier transform spectrometer (FTS) observations at the McMath Telescope to show that the absorption lines between  500 to 560 nm 
exhibited a 1.4\% decrease in {   amplitude} and 0.8\% decrease in equivalent width, at maximum compared with minimum of cycle 21. Interestingly,  \cite{livingstonetal2007} could not confirm this finding  in the next solar cycle for the unblended strong Fe I and the Na I D lines. This is, probably, a good indication of how challenging such observations are \citep[see also discussion in][]{penzaetal2006}. Instead,  \cite{livingstonetal2007}  reported cyclic variation in the Mn I 539.4 nm line and the CN 388.3 nm bandhead.  

Efforts have also been made to estimate the effect of  the Fraunhofer lines on the SSI variability with models. For example, \cite{sat_spectra} used spectra computed using the probability distribution functions and 
an estimate of  spot and facular disc area coverages during the solar maximum to show that spectral lines are the major contributor to the TSI variations in the 11-year activity cycle timescale rather than the continuum. In this study we follow up on the idea of \cite{sat_spectra} and employ the SATIRE \citep[Spectral And Total Irradiance Reconstruction, see][]{fliggeetal2000, krivovaetal2003} model to  assess the relative role of continuum and spectral lines contributions to the SSI variability. 
The availability of state-of-the-art measurements of solar disc area coverage of magnetic features and radiative transfer calculations of their spectra, that now account for the effects of non Local Thermodynamical Equilibrium (non-LTE), allows us to define the contribution of spectral lines to the irradiance variability over the entire solar spectrum (including the UV) and on different timescales of variability.


In Sect.~\ref{sect:model}  we describe our model. In Sect.~\ref{sect:contr} we show how the Fraunhofer lines affect the brightness contrasts between the different magnetic features and {   quiet regions} on the solar surface. In Sect.~\ref{sect:scales} we define the spectral profiles of the irradiance variability on the solar rotational, 11-year activity cycle, and centennial timescales. In Sect.~\ref{sect:lines_effect} we show how these three profiles look if we first exclude molecular lines from our calculations and then all Fraunhofer lines. The main results are summarised in Sect.~\ref{sect:conc}. 



\section{The model}\label{sect:model}
We base our calculations on the SATIRE model. 
Its branch SATIRE-S \citep[with ``S'' standing for the satellite era, see][]{SATIRE} has been  updated and refined over the recent years \citep[][]{balletal2012, balletal2014, yeoetal2014} and it currently replicates over 92\% of the observed TSI variability over the entire period of spaceborne observations.

In SATIRE magnetic features observed on the solar disc are divided into three classes: sunspot umbrae, sunspot penumbrae, {   and the combined faculae and the network.}  The part of the solar disc not covered by these magnetic features is attributed to the quiet Sun. The brightness of each component (i.e. one of the three magnetic feature classes or the quiet Sun) is assumed to be time-invariant, but depends on the wavelength and position on the visible solar disc. The solar irradiance is calculated by weighting  the spectra of the individual components with corresponding disc area coverages, i.e.:

\begin{equation}
S(t,\lambda)=\sum\limits_k \iint\limits_{\mathit{solar \,\, disc}} I_k \left ( \vec{\lambda, r}  \right ) \alpha_k (t, \vec{r}) \, d \Omega,
\label{eq:irr}
\end{equation}
where the summation is done over the three SATIRE classes of magnetic features and the quiet Sun.  For each of the components  the integration is performed over the visible solar disc. Here $I_k \left ( \vec{\lambda, r}  \right ) $  is the emergent intensity from the component $k$ at the wavelength $\lambda$ along the direction $\vec{r}$. The functions $\alpha_k (t, \vec{r})$ are the fractional coverages of the solar disc by the component $k$ along the direction $\vec{r}$, so that  $\alpha_k (t, \vec{r}) \, d \Omega$ gives the elementary solid angle covered by the component $k$.

The two main building blocks of our model are the spectra $I_k \left ( \vec{\lambda, r}  \right ) $ and the fractional coverages  $\alpha_k (t, \vec{r})$ of the quiet Sun and magnetic features.
We utilise the fractional coverages  deduced by  \cite{balletal2012} from the full-disc continuum images and magnetograms obtained by the Michelson Doppler Imager onboard the Solar and Heliospheric Observatory  \citep[SOHO/MDI;][]{scherreretal1995}. \cite{balletal2012} successfully used these fractional coverages  to reproduce  96\% of TSI variations during cycle 23 \citep[represented by the PMOD composite by ][]{ PMOD_comp}.  Following \cite{balletal2011,balletal2012}  we replace the integration in Eq.~(\ref{eq:irr}) with the summation over magnetogram pixels. Then the  $\alpha_k (t, \vec{r}) $ is substituted by $\alpha_{kij} (t)$ which represents the coverage of the pixel with abscissa $i$  and ordinate $j$   by the component $k$.The $\alpha_{kij} (t)$ values lie between 0 and 1 for the quiet Sun and faculae, and either 1 or 0 for umbra and penumbra. 

The brightness spectra of magnetic features and the quiet Sun $I_k \left ( \vec{\lambda, r}  \right ) $ are calculated with Non local thermodynamic Equilibrium Spectral SYnthesis Code (NESSY, Tagirov 2015, private comm.),  which is a further development of the COde for Solar Irradiance \citep[COSI,][]{shapiroetal2010}.  This is in contrast to previous SATIRE publications, which have been based on spectra calculated with the ATLAS9 code by \cite{kurucz1992} and \cite{ATLAS9_CK}. NESSY simultaneously solves the statistical equilibrium equations for the elements from hydrogen to zinc, taking the coupling between level populations of different elements via the electron concentration and radiation field into account.
 The employment of NESSY allows us to recalculate  the  $I_k \left ( \vec{\lambda, r}  \right ) $ spectra over the entire frequency domain  (in particular, including the UV, see Sect.~\ref{sect:scales}) with and without molecular/Fraunhofer lines.


We adopt the 1D temperature and  density structures of the quiet Sun, faculae, sunspot umbra, and sunspot penumbra  from Models C, P, S by \cite{fontenlaetal1999}, and Model R by \cite{fontenlaetal2006}, respectively.  The spectra emergent from magnetic features are calculated neglecting the Zeeman splitting. The populations of atomic and molecular levels, as well as the electron concentration are self-consistently calculated with NESSY. The atomic linelist is compiled from the ``long'' linelist provided by Kurucz (2006, private comm.) and the Vienna Atomic Line Database \citep[VALD,][]{kupkaetal1999,kupkaetal2000}. The molecular linelist is complied from the list by \cite{kurucz1993} and the Solar Radiation Physical Modeling database \citep[SRPM, see e.g.][]{fontenlaetal2011}. The strongest vibrational bands of the CN violet system and CH G-band are calculated  using the molecular constants by \cite{krupp1974,knowlesetal1988, wallaceetal1999}  \citep[see also][and references therein for more details]{shapiroetal2010, shapiro_CN}. 

The full disk brightness spectrum of the quiet Sun calculated with such a setup is in good agreement with various spaceborne measurements of the solar spectrum around activity minimum conditions \citep[see][for a detailed intercomparison]{shapiroetal2010,G_limb, IR1, IR3}. The centre-to-limb variations in the quiet Sun brightness agree well with measurements in the UV  \citep{LYRA_ECL} and visible (Tagirov 2015, private comm.) spectral domains. For the experiments presented below we use NESSY spectra calculated with the full molecular and atomic linelist, spectra produced excluding molecular lines, and finally spectra calculated excluding all Fraunhofer lines, i.e. when the total opacity is given only by the free-free and bound-free processes.




\section{Effect of the Fraunhofer lines on contrasts of solar magnetic features}\label{sect:contr}
\begin{figure*}
\resizebox{!}{0.4\vsize}{\includegraphics{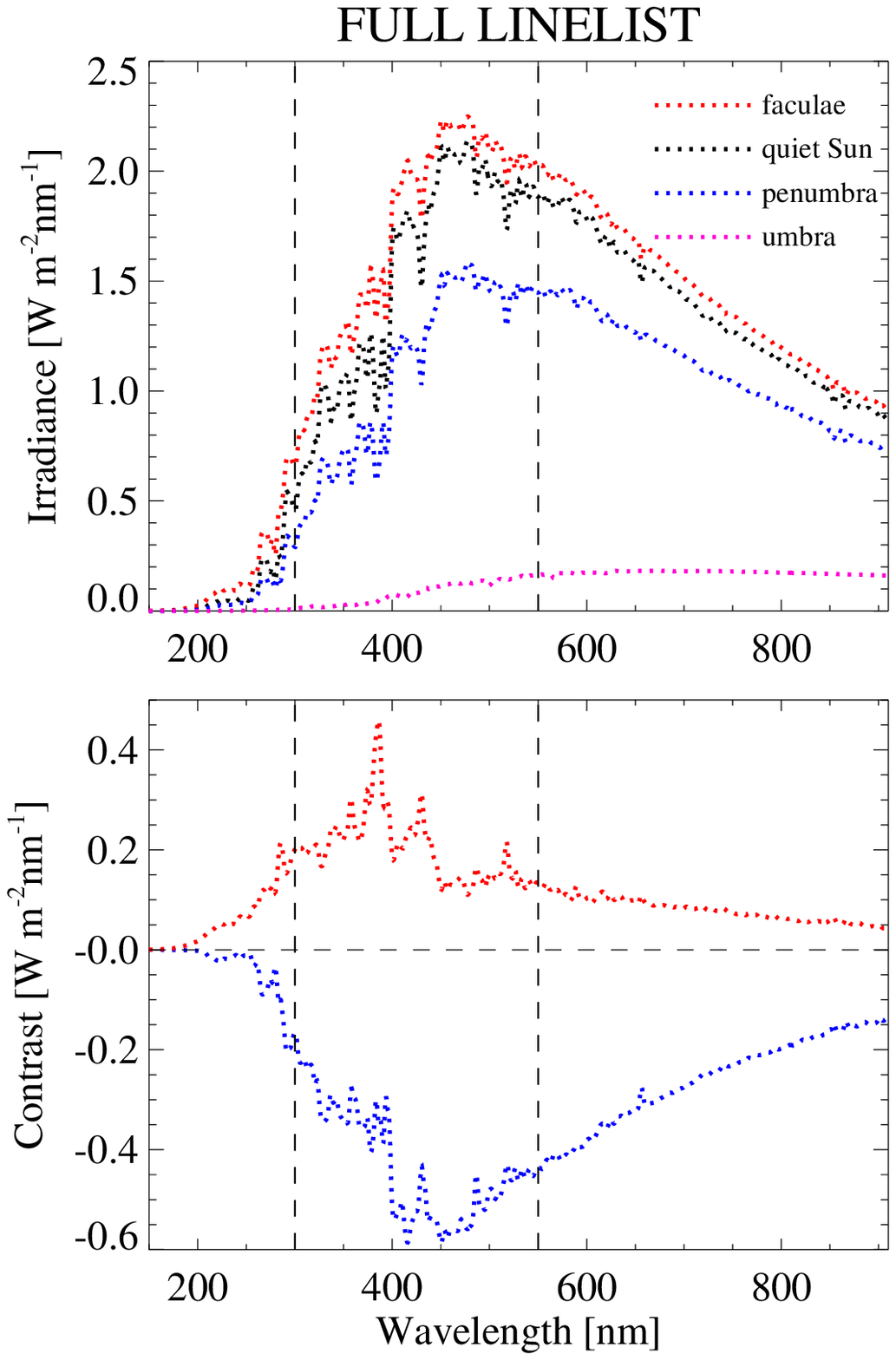}}
\resizebox{!}{0.4\vsize}{\includegraphics{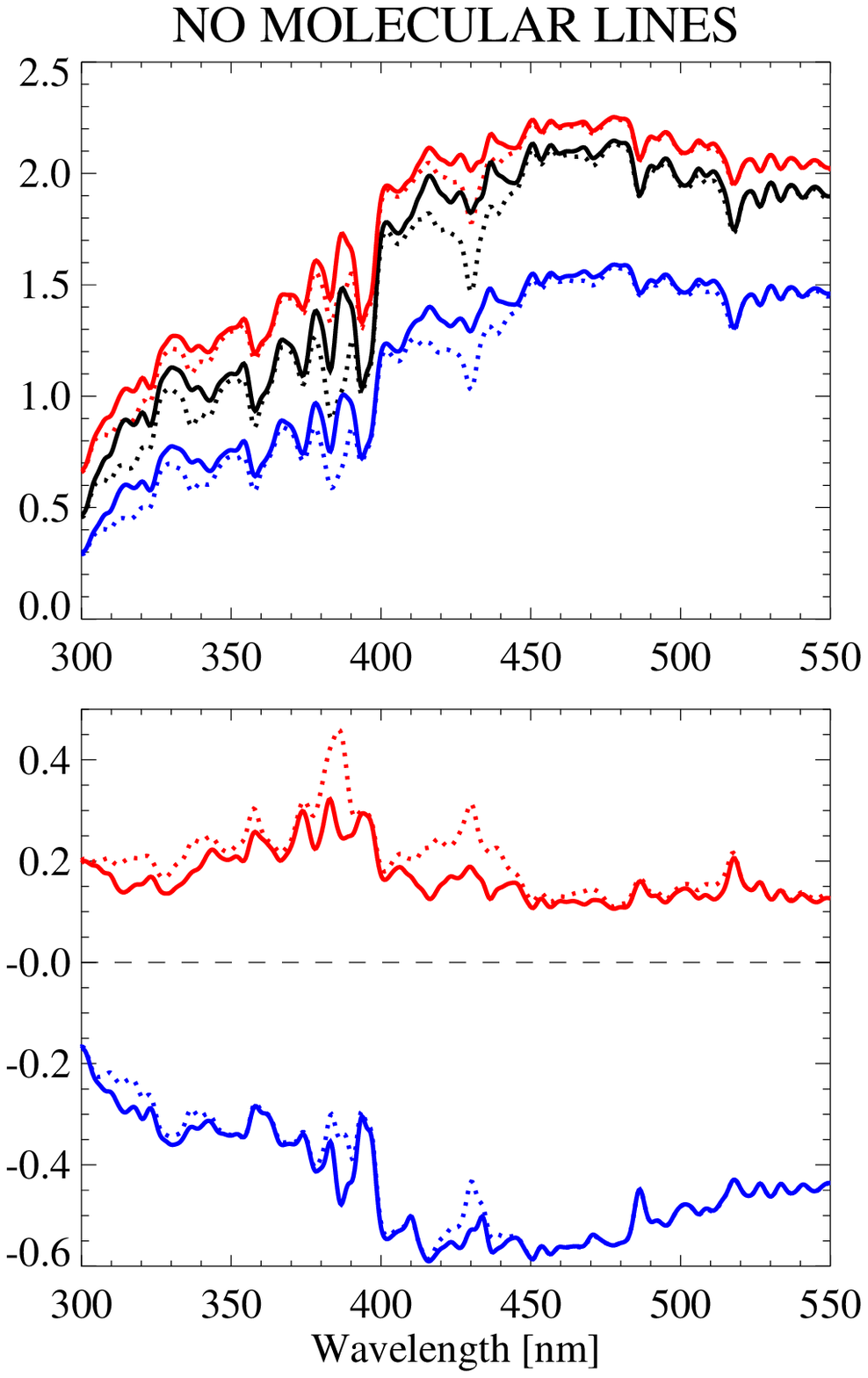}}
\resizebox{!}{0.4\vsize}{\includegraphics{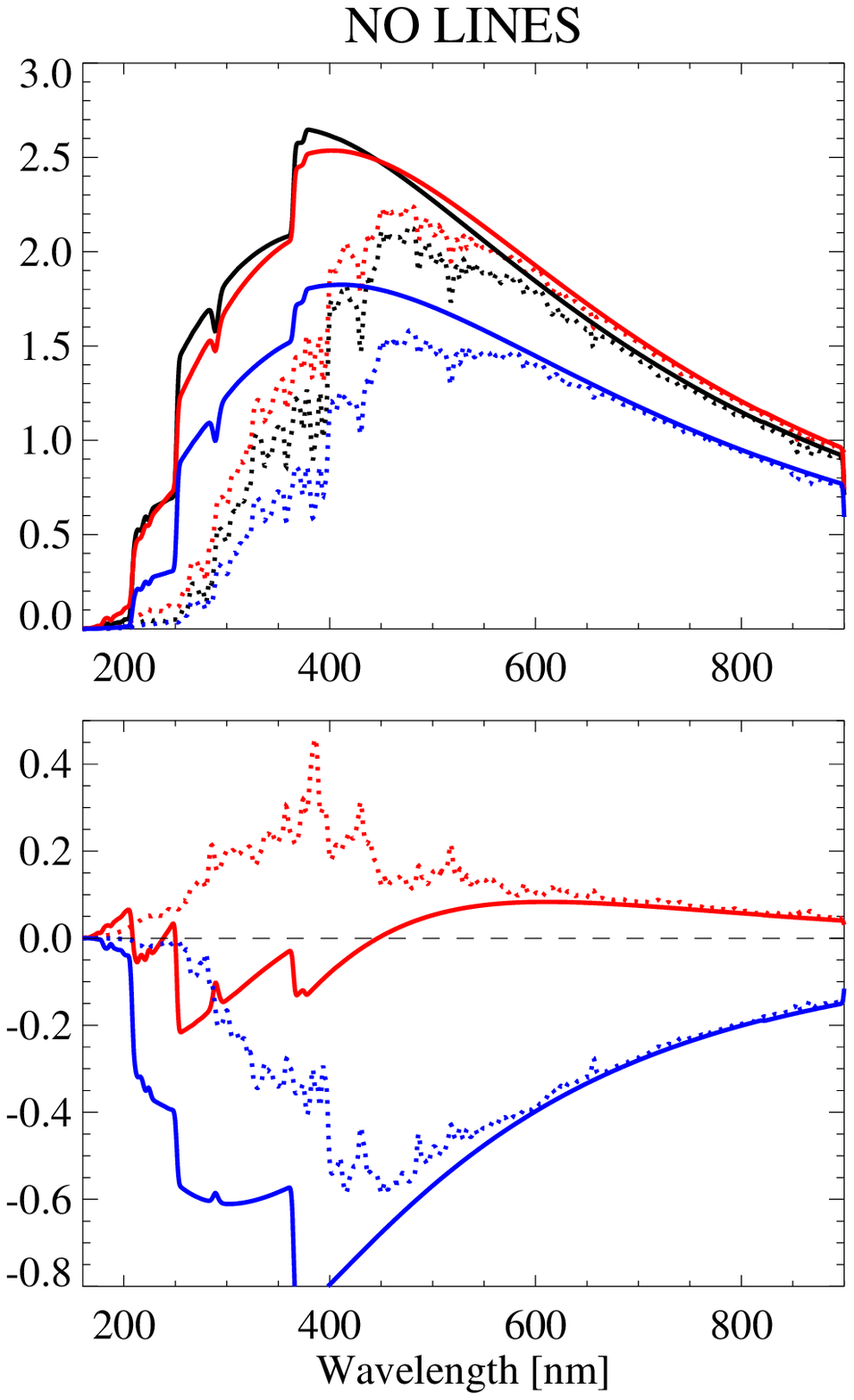}}


\caption{Upper panels: the brightness spectra of the quiet Sun (black curves), faculae (red curves), spot penumbra (blue curves) and umbra (magenta curve, only plotted in the upper left panel).  Lower panels: the facular (red curves) and penumbral (blue curves) brightness contrasts with respect to the quiet Sun. The dotted curves are calculated taking the spectral lines (full linelist) into account. The solid curves are calculated by putting the opacity in the molecular (middle panels) and in all spectral lines (right panels) to zero. To facilitate the comparison between different spectra and contrasts the dotted curves are plotted in all three panels. The two vertical dotted lines in the left panel constrain the spectral interval shown in the middle panels. }
\label{fig:contr}
\end{figure*}
In the upper left panel of Fig.~\ref{fig:contr} we plot the spectral fluxes at 1 AU (hereafter referred as brightness spectra or just brightness) emergent from the quiet Sun, faculae, spot umbrae and penumbrae calculated as discussed in Sect.~\ref{sect:model}. For illustration purposes the brightness of each type of magnetic feature is plotted assuming that it fully covers the visible solar disc. 

With the exception of the umbral brightness, the spectral profiles of the brightness of different magnetic  components are strongly affected by numerous absorption features.  The umbral brightness is significantly lower than brightnesses of all other magnetic components considered in this study and the spectral lines have only marginal contribution to the umbral brightness contrast with respect to the quiet Sun.  Hence the effect of spectral lines on the umbral spectrum will not be discussed in detail, though it will be taken into account when computing irradiance variability.

In the lower left panel of Fig.~\ref{fig:contr} we present  facular and penumbral brightness contrasts with respect to the quiet Sun. They are calculated by subtracting the brightness of the quiet Sun from the facular and umbral brightness.  The overall profiles of the contrasts are to some extent reminiscent of the solar spectrum but skewed towards the UV because  {   at short wavelengths  (i.e. when the Planck function can be approximated by the Wien approximation) the derivative of the Planck function with respect to the temperature (which defines the contrast) is proportional to the Planck function divided by the wavelength}.  

To illustrate the contribution of spectral lines to the facular and penumbral brightness contrasts  we also show brightness spectra and contrasts calculated by putting the opacity of molecular lines to zero (middle panels of  Fig.~\ref{fig:contr}) as well as putting opacity in all Fraunhofer lines to zero (i.e. producing purely continuum spectra and contrasts, see right panels of Fig.~\ref{fig:contr}). We discuss these plots in Sects.~\ref{sect:contr.mol} and \ref{sect:contr.atom}, respectively.

\subsection{Molecular lines}\label{sect:contr.mol}
The middle panels of Fig.~\ref{fig:contr} illustrate that the elimination of molecular lines from the radiative transfer calculations has a strong effect on the brightness spectra and the contrasts in the 300 -- 450 nm spectral domain. The strongest peak in the facular contrast is linked to the CN violet system ($B {}^{2} \Sigma - X {}^{2} \Sigma$ transitions)  between 380 nm and 390 nm, while the second strongest peak results from the CH G-band ($A ^2 \Delta - X ^2 \Pi$ transitions) around 430 nm. These molecular systems, being also prominent features of the solar spectrum, almost double the facular contrast at the corresponding wavelengths.  
The brightness contrasts and spectra between 300 and 350 nm are also noticeably affected by the CN, NH, and OH bands.

The amplification of facular contrast by molecular lines is due to the strong sensitivity of molecular concentrations to temperature changes. Molecular concentrations are lower in faculae which are generally warmer than the quiet Sun surroundings. Consequently, the molecular lines in the facular spectrum are weaker than in the spectrum of the quiet Sun, which leads to the enhancement of the contrast. Such an interpretation is in line with \cite{steiner1, Almeidaetal2001}. Similarly to this study, they employed 1D models of faculae and quiet Sun to explain the appearance of the G-band bright points, observed on G-band filtergrams, by the weakening of CH lines in hot flux tubes. 
Similar conclusion was {   later} drawn by  \cite{mol_BP}  who  employed more realistic 3D MHD simulations \citep[see also][who studied the CN bright points]{CN_BP}.

The CN violet system and CH G-band induce a comparable  amplification of the facular brightness contrast even though the CH G-band is more pronounced in the solar spectrum. 
This is partly caused by the relatively high dissociation energy of the CN molecule (7.72 eV), which is larger than that of CH (4.25 eV). As a consequence the CN concentration is more sensitive to temperature changes than the CH concentration \citep[see e.g. the discussion of the chemical equilibrium calculations in][]{berdyuginaetal2003}.  

The CN violet system and CH G-band are less pronounced in the penumbral spectrum than in the spectrum of the quiet Sun, which is probably associated with a small gradient of photospheric temperature 
in the \cite{fontenlaetal2006} penumbral model (see their Fig. 6). As a result the penumbral brightness contrast drops in the CH G-band and is only marginally affected by the CN violet system. 




\subsection{Fraunhofer lines}\label{sect:contr.atom}
The right panels of  Fig.~\ref{fig:contr} illustrate that the elimination of all Fraunhofer lines from the radiative transfer calculations has a profound effect on both the brightness spectra and the contrasts below about 600 nm. The spectral lines are a dominant source of opacity in the UV and they significantly contribute to the total opacity in the visible spectral domain. By shifting the formation height of the radiation to higher and colder photospheric layers, the Fraunhofer lines significantly reduce the brightness of the quiet Sun, faculae, and penumbrae (see upper right panel of Fig.~\ref{fig:contr}).  The salient features of the continuum spectra are the Balmer, \ion{Mg}{I}, and \ion{Al}{I} ionisation edges at 364.5 nm, 251.2 nm, and 207.8 nm, respectively. The feature around 290 nm is attributed by \cite{Hab_op} to a resonance in the \ion{Mg}{I} 3 photoionisation cross section. While the \ion{Mg}{I} and \ion{Al}{I} edges can still be recognised in the ``real'' (i.e. calculated with the full linelist, see dotted lines in  Fig.~\ref{fig:contr}) solar spectrum, the Balmer edge is completely hidden in the multitude of spectral lines. 

The difference between the quiet sun and faculae continua is small only, the former slightly brighter in the middle and near UV {   (200--300 nm and 300--400 nm, respectively)}, the latter slightly brighter shortward of 200 nm and longward of 400 nm. The penumbra continuum is well below the other two, as can be expected from its effective temperature that is about 250K lower than that of the quiet sun. 

The Fraunhofer lines substantially increase the facular contrast and decrease the absolute value of the penumbral contrast.  This is linked to the temperature structure of the 1D atmospheric models  (see Sect.~\ref{sect:model}): the temperature difference between the quiet Sun and faculae decreases, while the temperature difference between the quiet Sun and penumbra increases, towards deeper photospheric layers \citep[see Fig. 6 from][]{fontenlaetal2006}, where the continuum intensity is formed. The effect of Fraunhofer lines on the brightness spectra and contrasts gets smaller in the infrared where the relative contribution of lines to the total opacity diminishes. 

According to our calculations the facular contrast in the continuum becomes negative in the near and middle UV.  The continuum irradiance at these wavelengths comes from the deepest photospheric layers \citep[see also][]{Ayres1989}. Indeed, at shorter wavelengths the photoionisation opacity from metals {   increases}, while at longer wavelengths the {   opacity from negative hydrogen ions get higher} \citep[see Fig. 4.2 from ][]{mihalas1978}. In deep photospheric layers faculae are cooler than the quiet Sun according to the 1D atmospheric structures employed here, which results in the inversion of the contrast. A similar inversion in the continuum contrast was recently obtained by \cite{SIM_MHD} who based their calculations on 3D MHD simulations calculated by \cite{fabbian2012} using STAGGER code \citep{STAGGER1}.

The negative full-disc facular continuum brightness contrasts in the UV obtained from our calculations are qualitatively consistent with the observations of negative continuum brightness contrasts of faculae about disc centre in the visible part of the solar spectrum \citep[][and references therein]{yeoetal2013} where the radiation also emanates from the deep photosphere. We note that the contribution functions \citep[see e.g.][p.151]{gray1992} for the full-disk UV and disc centre visible continuum radiation are comparable because the formation height of the continuum radiation decreases with cosine of the heliocentric angle but increases from the UV to visible. At the same time we are aware that while 1D models are designed to replicate many features of brightness spectra and contrasts \citep[see e.g.][]{sat_spectra, fontenlaetal2011} they miss the real physics of centre-to-limb variations in the facular contrast   \citep{steiner2}.

\section{Spectral irradiance variability on different timescales}\label{sect:scales}
\begin{figure}
\resizebox{\hsize}{!}{\includegraphics{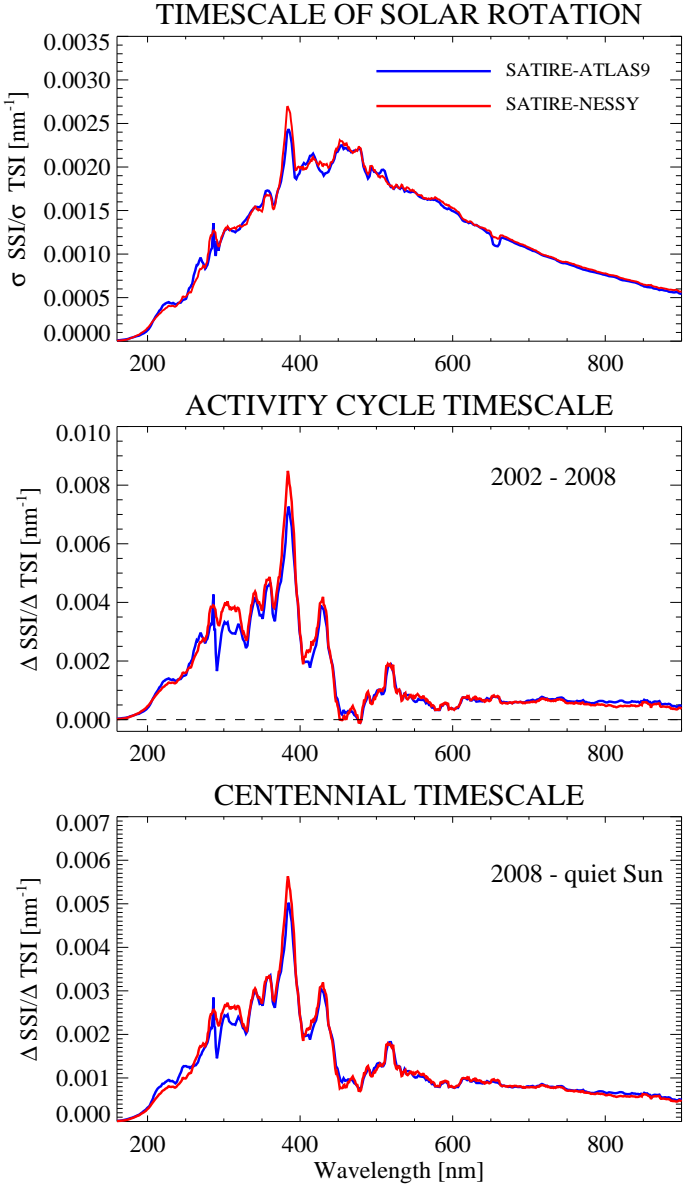}}
\caption{The spectral profiles of the irradiance variability calculated on the solar rotational (upper panel), activity cycle (middle panel), and centennial (lower panel) timescales. Plotted are the values calculated employing  ``SATIRE-ATLAS9'' SSI and TSI time series (blue curves) and values calculated employing the ``SATIRE-NESSY''  SSI and TSI time series (red curves). All spectral profiles are smoothed by applying a 10 nm running mean.}
\label{fig:prof}
\end{figure}
The most accurate and uninterrupted records of solar irradiance variability available so far are the TSI records \citep[see e.g. the most recent publications by][and references therein]{koppetal2012, Claus_rev, schmutzetal2013, koppetal2014}.  Nevertheless more information than just TSI is needed for both climate simulations and the solar-stellar comparison. Climate simulations are very sensitive to the variability of the UV irradiance, which is the main driver in the top-down mechanism for the influence of the solar irradiance variability on climate \citep{haigh1994, top-down}. Likewise, most stellar photometric records used for solar-stellar comparison are obtained with broad-band filters sensitive to the visible part of the spectrum \citep{lockwoodetal2007,halletal2009,basrietal2013}. It is, therefore, important to know the spectral profiles of the solar irradiance variability on different timescales. In this Section we explain how these spectral profiles are linked to the brightness contrast discussed in Sect.~\ref{sect:contr}. Then in  Sect.~\ref{sect:lines_effect} we show how these profiles are affected by the Fraunhofer lines.

We define the spectral profile of the irradiance variability on the timescale of solar rotation as:
\begin{equation}
\frac{\sigma  {\rm SSI}}{\sigma {\rm TSI}} (\lambda)=\frac{{\rm rms} \left ( {\rm SSI}(\lambda,t)\, - \left <{\rm SSI}(\lambda,t) \right >_{81}  \right )}{{\rm rms} \left (  {\rm TSI}(t)\, - \left < {\rm TSI}(t) \right >_{81} \right )},
\label{eq:rot}
\end{equation}
where ``rms'' stands for the root mean square. In our study both TSI and SSI time series are calculated with daily cadence so that the averaging in Eq.~\ref{eq:rot} represents the running mean over 81 days.

In the upper panel of Fig.~\ref{fig:prof} we plot ${\sigma  {\rm SSI}}/{\sigma {\rm TSI}}$ values as functions of wavelength calculated with SATIRE-S TSI and SSI time series from \cite{balletal2012, balletal2014}, hereafter ``SATIRE-ATLAS9'' time series, since they are based on the spectra of individual magnetic features and the quiet Sun calculated with the ATLAS9 code. These time series cover the period from February 19, 1999 till October 2, 2010 and are based on the fractional coverages of magnetic features deduced from the SOHO/MDI data. We also plot  ${\sigma  {\rm SSI}}/{\sigma {\rm TSI}}$ values calculated with the model described in Sect.~\ref{sect:model},  i.e. employing the same SOHO/MDI   fractional coverages of magnetic features as \cite{balletal2012, balletal2014} but convolving them with the NESSY spectra instead of the ATLAS9 spectra (see Sect.~\ref{sect:model}). These ``SATIRE-NESSY'' time series are calculated for the same period as the ``SATIRE-ATLAS9'' time series.

In the middle panel of  Fig.~\ref{fig:prof}  we plot  ``SATIRE-NESSY'' and ``SATIRE-ATLAS9''  spectral profiles of the irradiance variability on the 11-year activity cycle timescale. They are defined as 
\begin{equation}
\frac{\Delta  {\rm SSI}}{\Delta {\rm TSI}} (\lambda)= \frac{ \left <{\rm SSI}(\lambda,t)  \right  > _{2002} - \left <{\rm SSI}(\lambda,t)  \right  >_{2008}}{ \left <{ \rm TSI}(t)  \right  > _{2002} -  \left <{\rm TSI}(t)  \right  >_{2008}},
\label{eq:11}
\end{equation} 
where annual averaging is performed over {   the calendar years of} high  (year 2002) and low (year 2008) solar activity.

The irradiance variability on a centennial timescale is often described as a secular change between solar minima conditions \citep[see e.g. the discussion of TSI variability in][]{Claus2009}. The current solar irradiance records are too short and uncertain to unambiguously reveal and quantify secular changes. Consequently, the magnitude and even specific physical mechanisms responsible for the centennial SSI variability  are heavily debated \citep[see][and references therein]{MPS_AA}. 
According to SATIRE most of the SSI changes between activity minima are caused by the varying contribution from the network component \citep{krivova_rec2010}.  The contribution of this component does not drop to  zero even at activity cycle minima and is responsible for the secular trend in solar irradiance, e.g. for the change in irradiance between the 2008 activity minimum and the Maunder minimum  \citep[however, see also][]{schrijveretal2011}.

Along these lines we define the spectral profile of the irradiance variability on {   the} centennial timescale as
\begin{equation}
\frac{\Delta  {\rm SSI}}{\Delta {\rm TSI}} (\lambda)= \frac{  \left  <{\rm SSI}(\lambda, t) \right > _{2008} - {\rm SSI}_{\rm quiet}(\lambda)}{  \left  <{\rm TSI}(t) \right > _{2008} - {\rm TSI}_{\rm quiet}}, 
\label{eq:lt}
\end{equation}
where $ {\rm SSI}_{\rm quiet}(\lambda)$ and ${\rm TSI}_{\rm quiet}$ are spectral and total solar irradiance, respectively, as they would be measured if the visible part of the solar disc did not  contain any active features (i.e. was completely covered by the quiet Sun). {   We emphasise that our definition of the centennial variability refers to the irradiance changes between solar activity minima rather than averaged over the activity cycle as often done in the literature.} The spectral profiles on {   the} centennial timescale calculated with the ``SATIRE-NESSY'' and ``SATIRE-ATLAS9'' time series are plotted in the lowest panel of Fig.~\ref{fig:prof}.

{   In addition to strong concentrated magnetic fields, considered in SATIRE, there is also weak turbulent magnetic field on the solar surface \citep[see reviews by][]{deWijnetal2009, stenflo_rev}. Its effect on the rotational and 11-year irradiance variability is believed to be small \citep[cf.][]{MPS_AA} but it is presently unclear whether this field can contribute to  the irradiance variability on the centennial timescale \citep[see e.g. discussion in][and references therein]{analysis}. The ambiguity associated with this effect  is an additional source of uncertainty in  our estimate of the contributions of molecular and atomic lines to irradiance variability on the centennial timescale. }

We note that the ATLAS9 code is based on the LTE assumption, which is not applicable in the UV. As a consequence \cite{balletal2014} corrected the SSI values between 115 and 270 nm based on the empirical method by \cite{krivova_UV}. In contrast to ATLAS9, NESSY  takes non-LTE effects into account and thus no additional correction of the ``SATIRE-NESSY'' time series is needed.  Despite this difference in methods,  the spectral profiles calculated with both time series are very close to each other over the entire spectral range shown in Fig.~\ref{fig:prof}, also including the UV. 
We construe this as an independent support to both  the \cite{krivova_UV} method and the reliability of the ``SATIRE-NESSY'' time series. 

The  ``SATIRE-ATLAS9'' SSI and TSI time series have been demonstrated to be consistent with observations from multiple sources \citep[see][and references therein]{balletal2014,yeoetal2014}. Since the differences between ``SATIRE-NESSY'' and ``SATIRE-ATLAS9'' spectral profiles are not essential for the purposes of this study we refrain from comparing the ``SATIRE-NESSY'' time series to observational data here, which is beyond the scope of this paper.

Figure~\ref{fig:prof} indicates that the spectral profile of the irradiance variability depends significantly on the considered timescale.  The SSI variability on the 11-year activity cycle timescale (middle panel of Fig.~\ref{fig:prof}) is brought about by the competition between bright faculae and network on the one hand and dark sunspots on the other hand. The 11-year TSI variability as well as the 11-year variability in the UV and visible are faculae-dominated and are in phase with the solar cycle \citep[however, see also][]{harderetal2009}, so that the $\Delta  {\rm SSI}/\Delta {\rm TSI}$ values plotted in the middle panel of Fig.~\ref{fig:prof} are positive with the exception of a minor feature around 470 nm. The facular contribution  decreases with wavelengths faster than the spot contribution (in particular, the umbral component, see left upper panel of Fig.~\ref{fig:contr}) so that starting from 600 nm both components almost cancel each other \citep[while starting from  1-2 $\rm {\mu m}$ spots overweight the faculae and SSI varies out-of-phase with solar cycle, see e.g. Fig.~7 from][]{TOSCA2012}. As a result the dominant part of the irradiance variability over the 11-year activity cycle originates in the 250 -- 450 nm spectral domain, which is especially affected by the molecular lines (see middle panels of Fig.~\ref{fig:contr}).

The SSI variability on the timescale of solar rotation is mainly caused by transits of magnetic features across the visible solar disc as the Sun rotates. The spot and facular contributions to the irradiance rarely cancel each other {\it at any specific moment of time}. 
Hence, in contrast to the case of the 11-year variability, the amplitude of the irradiance variability on the rotational timescale does not drop in the visible part of the spectrum and is still significant in the red and IR spectral domains (where it is mainly associated with umbral contributions). Consequently  the role of the wavelengths affected by the molecular lines is significantly smaller than in the case of the 11-year variability.


 


In our model the spectral profile of the irradiance variability on {   the} centennial timescale is directly given by the spectral profile of the facular brightness contrast (because the same atmospheric model is used for faculae and network, see Sect.~\ref{sect:model}). 
In the infrared the centennial SSI variability is larger than the 11-year SSI variability (since there is no competing effect from spots) but is substantially smaller than the variability on the time scale of solar rotation.



\section{Effect of the Fraunhofer lines on spectral irradiance variability}\label{sect:lines_effect}
\begin{figure}
\resizebox{\hsize}{!}{\includegraphics{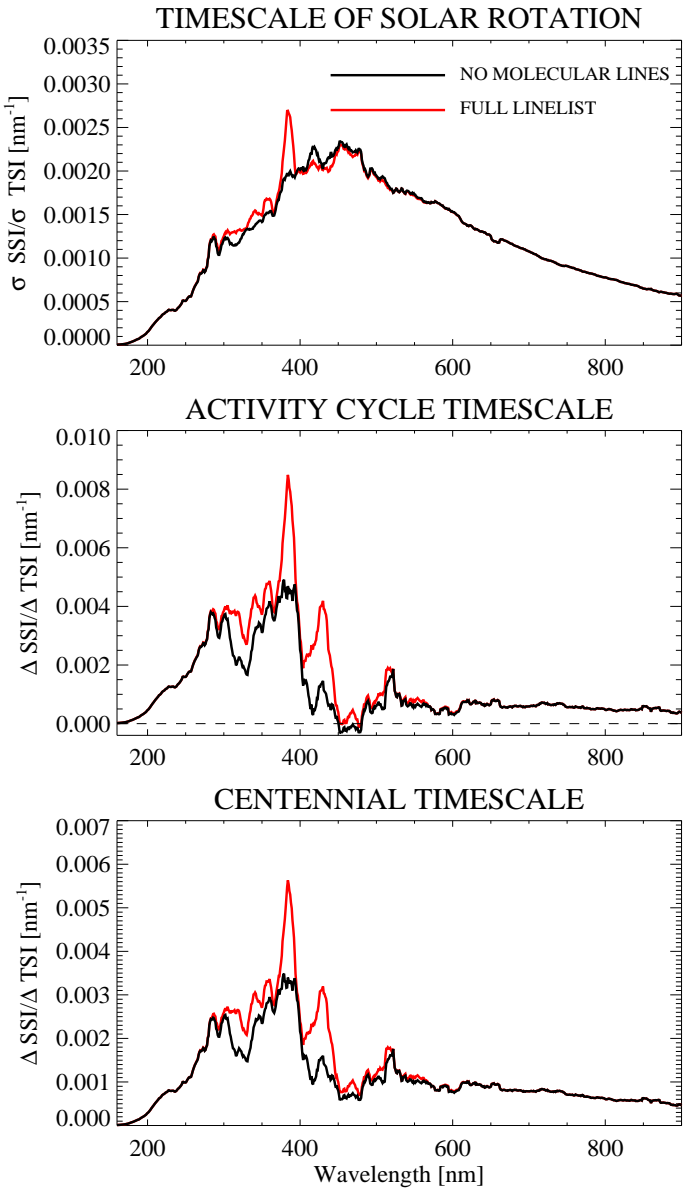}}
\caption{The spectral profiles of the irradiance variability calculated on the solar rotational (upper panel),  11-year activity cycle (middle panel), and centennial (lower panel) timescales. Plotted are the values calculated with ``SATIRE-NESSY'' model taking the full linelist into account (red curves) as well as putting the opacity in molecular lines to zero (black curves).}
\label{fig:prof_mol}
\end{figure}
Despite the difference in overall shape, all three spectral profiles of the irradiance variability, introduced in Sect.~\ref{sect:scales}, have similar spectral features which are mostly attributed to different molecular bands and strong atomic lines. To pinpoint the contributions of variations in molecular and atomic lines to these profiles  we calculated intensity spectra $I_k \left ( \vec{\lambda, r}  \right )$   by first putting the opacity in molecular lines to zero and then the opacity in all Fraunhofer lines to zero. For each of the computations Eq.~(\ref{eq:irr}) was then used to produce  ``SATIRE-NESSY $_{no\,\,molecules}$'' and  ``SATIRE-NESSY $_{no\,\,atoms}$''  SSI time series, respectively. These SSI time series were then substituted  into Eqs.~(\ref{eq:rot})--(\ref{eq:lt}) to recalculate the spectral profiles of the irradiance variability as they would be measured without contributions from molecular/Fraunhofer lines.  To maintain a consistent normalisation of the spectral profiles, the TSI time series used in  Eqs.~(\ref{eq:rot})--(\ref{eq:lt}) was calculated  taking the molecular and atomic lines into account, i.e. is always the same.


\subsection{Molecular lines}
In Fig.~\ref{fig:prof_mol} we present a comparison of the spectral profiles of the SSI variability calculated employing the ``SATIRE-NESSY'' (produced with the full linelist)  and ``SATIRE-NESSY $_{no\,\,molecules}$'' SSI time series. In line with the discussion in Sect.~\ref{sect:contr.mol}, molecular lines mainly affect the near UV, violet, and blue spectral domains. 
They almost double the SSI variability on the 11-year and centennial timescales between 300 and 450 nm. As mentioned in Sect.~\ref{sect:scales}  this spectral domain  also happens to be the major contributor to the TSI variability on the 11-year timescale. By integrating the spectral profiles of the centennial and 11-year variability over the wavelengths one can estimate the contribution of the molecular lines to the TSI variability.  Our calculations indicate that 23\% of the TSI variability on the 11-year timescale is attributed to molecular lines. On {   the} centennial timescale (bottom panel of Fig.~\ref{fig:prof_mol}) the  relative contribution of the 300--450 nm spectral domain to the TSI variability is smaller than in the case of the 11-year variability. Consequently  the contribution of molecular lines to the TSI variability on {   the} centennial timescale is also slightly lower, being equal to 15\%.

In contrast to the 11-year and centennial variabilities, molecular lines do not play an essential role in the variability on the timescale of solar rotation. This is mainly because of the spot contribution which keeps the rotational variability high even in the absence of molecular lines.

Molecular lines are also responsible for the strongest peak in the SSI variability on all three timescales considered in this study, namely the 380--390 nm peak which is linked to the CN violet system. The second strongest peak in the SSI variability on the 11-year and centennial timescales is associated with the CH G-band.

\subsection{Fraunhofer lines}
\begin{figure}
\resizebox{\hsize}{!}{\includegraphics{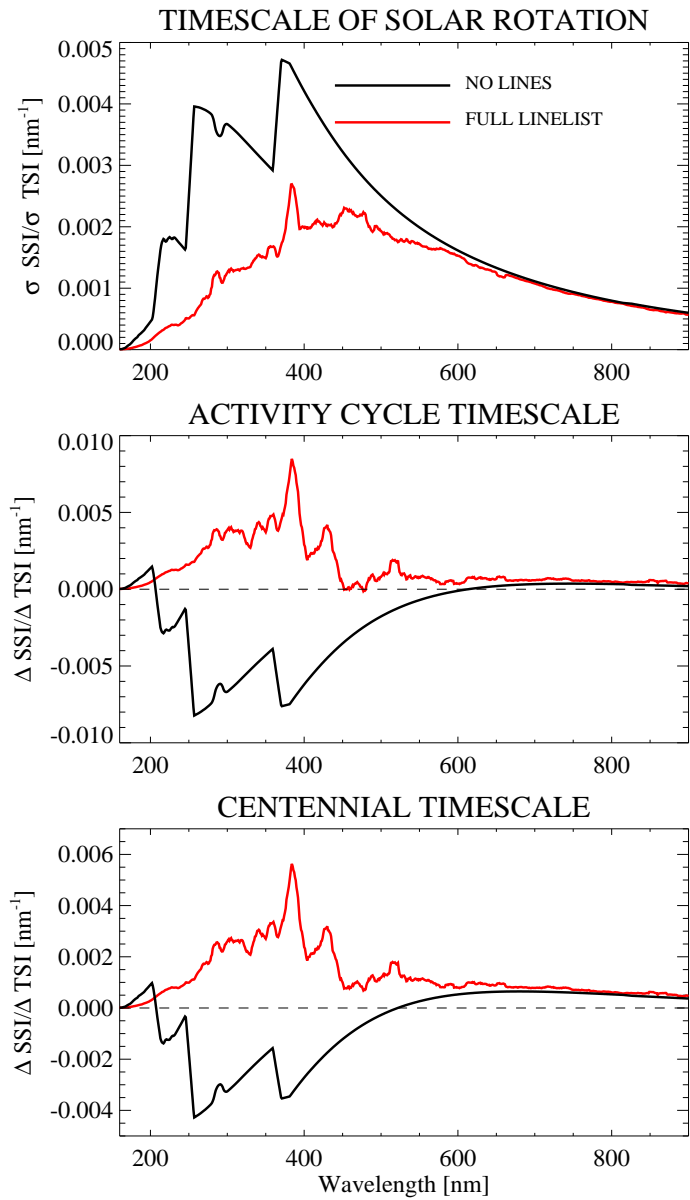}}
\caption{The same as Fig.~\ref{fig:prof_mol} but now the black curves represent the values calculated putting the opacity in all Fraunhofer lines to zero.}
\label{fig:prof_all}
\end{figure}

In Fig.~\ref{fig:prof_all} we compare the spectral profiles calculated employing the ``SATIRE-NESSY'' and ``SATIRE-NESSY $_{no\,\,atoms}$'' SSI time series.
The elimination of the Fraunhofer lines has a profound effect on all three spectral profiles up to 600 nm. The ionisation edges clearly visible in the brightness spectra and contrasts in the right panels of Fig.~\ref{fig:contr} are also prominent features of spectral profiles of continuum variability. 

Notably our calculations indicate that the continuum SSI variability in the UV and visible is in antiphase with solar activity  on the 11-year and centennial timescales. This is consistent with \cite{sat_spectra}, who also reported an antiphase continuum variability on the 11-year timescale for the 300--500 nm spectral domain \citep[cf. also][]{SIM_MHD}. 

Furthermore, since the 11-year and long-term TSI variability calculated without accounting for spectral lines is out-of-phase with the solar cycle, the contributions of Fraunhofer lines to the TSI variability on these timescales is larger than 100\%, being 250\% and 140\% for the 11-year and long-term timescales, respectively. By the same token the increase of the TSI at maximum of the activity cycle compared with its minimum is directly attributed to the variability in spectral lines. We note, however, that the exact behaviour of the continuum variability strongly depends on the temperature structures of the quiet Sun and magnetic features in the deepest photospheric layers. These layers provide a very small contribution to the emergent line spectra, so that it is difficult to reliably constrain  their temperature structures in 1D semi-empirical modelling, especially taking intro account the uncertainties in the measurements of the IR solar irradiance which emanates from the deep photospheric layers \citep{IR1,IR2,IR3}. Consequently our model does not allow us to unambiguously rule out an in-phase continuum variability. At the same time our calculations clearly indicate that the variations in spectral lines govern the TSI variability on all three considered timescales.

\section{Discussion and Conclusions}\label{sect:conc}
We have estimated the contribution of the Fraunhofer lines to the solar irradiance variability driven by the evolution of the surface magnetic field on the solar rotational, 11-year activity cycle, and centennial timescales. Our calculations indicate that the solar irradiance variability in the UV, violet, blue, and green spectral domains is fully controlled by the Fraunhofer lines.  The highest peak in {\it absolute} SSI variability on all considered timescales is associated with the CN violet system between 380 and 390 nm. Furthermore, the molecular lines strongly enhance the SSI variability on the 11-year and centennial timescales. For example, according to our model, almost a quarter of the TSI variability on the 11-year timescale originates in molecular lines.  

Such a substantial  contribution of molecular lines to the irradiance variability is due to the strong sensitivity of molecular concentrations to temperature changes and consequently an enhanced molecular depletion in hot facular regions compared to the colder surrounding regions of the quiet Sun. This is a well known effect, 
which leads to the appearance of bright points in molecular lines at the locations of magnetic flux concentrations \citep{steiner1, mol_BP} and makes molecular lines a sensitive tool for mapping photospheric temperatures and magnetic fields  \citep{berdyuginaetal2005}. However, to our knowledge, it is the first time that molecular lines are directly linked to the enhancement of the irradiance variability.

Some words of caution are needed here since our model is based on the NLTE calculations with static 1D atmospheric stratifications of the quiet Sun and magnetic features. Such calculations are still being actively developed and are getting highly sophisticated \citep[see e.g. discussion in][]{ruttenuitenbroek2012}. At the same time  they are deemed \citep[e.g.][]{koesterkeetal2008,1D_bad} to be inherently unreliable for diagnostics of solar atmospheric properties \citep[see however][]{Vitasetal2009}, unless differential techniques are used \citep{stenfloetal1998}.
The 1D models cannot directly account for the hot walls of flux tubes which to large extent determine the centre-to-limb variation in facular brightness \citep{steiner2}. Furthermore, molecular concentrations depend on the temperature in a non-linear way  \citep[see e.g.][]{berdyuginaetal2003} so that their response to horizontal temperature fluctuations  cannot be properly averaged out in 1D calculations. Also, for some molecules the assumption of instantaneous chemical equilibrium may fail \citep[][]{ramosetal2003}.

The temperature structures of 1D solar atmospheric models are specifically adjusted to satisfy as many observational constrains as possible. 
This makes 1D models quite successful at reproducing the observations relevant for calculations of the SSI variability \citep[see e.g.][and references therein]{sat_spectra, fontenlaetal2011}.  Additionally, despite their huge potential,  self-consistent 3D radiative transfer calculations of the entire solar spectrum based on the results of MHD simulations and capable of reproducing high-resolution observations are not yet readily available  \citep[see e.g.][]{nadine2011}.   Hence, despite the obvious limitations, the semi-empirical 1D calculations still remain at the forefront of irradiance modelling.  At the same time 3D MHD simulations have been gradually reaching a new level of realism and there are also intermediate approaches in between 1D and 3D radiative transfer \citep[][]{ayresetal2006,HS2012,HS2013}. It would be interesting to repeat the exercise presented in this paper when such calculations are applied to modelling of the solar irradiance variability. A study along these lines was recently performed by \cite{SIM_MHD}, who, however, only calculated the facular brightness at continuum wavelengths.

Our result indicates that proper calculations of opacity in the solar atmosphere are of high importance for modelling of the solar irradiance variability. The inclusion of Fraunhofer lines is absolutely crucial for reproducing variations in TSI and SSI at all timescales from the solar rotation period to centuries. For quantitatively accurate computations molecular lines also need to be taken into account, in particular CN and CH.




\begin{acknowledgements}
{   We thank Greg Kopp for careful reading and constructive and useful comments on the manuscript.} The research leading to this paper has received funding from the People Programme (Marie Curie Actions) of the European Union's Seventh Framework Programme (FP7/2007-2013) under REA grant agreement No. 624817. It also got financial support  from the BK21 plus program through the National Research Foundation (NRF) funded by the Ministry of Education of Korea. RVT acknowledges support through SNF grant 200020\_153301. 
\end{acknowledgements}

\bibliographystyle{aa}

\end{document}